\documentclass[%
reprint,
amsmath,amssymb,
aps,
]{revtex4}
\usepackage[utf8]{inputenc}
\usepackage{cancel}
\usepackage{graphicx}
\begin{document}
\title{Higher-order one-loop renormalization in the spinor sector of minimal LV extended QED}

\author{L. C. T. Brito}
\email{lcbrito@ufla.br}
\affiliation{Departamento de F\'{i}sica, Universidade Federal de Lavras, Caixa Postal 3037,
37200-000, Lavras, MG, Brasil}

\author{J. C. C. Felipe}
\email{jean.cfelipe@ufvjm.edu.br}
\affiliation{Instituto de Engenharia, Ci\^{e}ncia e Tecnologia, Universidade Federal dos Vales do Jequitinhonha e Mucuri, Avenida Um, no.  4050,  39447-814, Cidade Universit\'{a}ria,  Jana\'{u}ba,  MG, Brazil}

\author{J. R. Nascimento}
\email{jroberto@fisica.ufpb.br}
\affiliation{Departamento de F\'{i}sica, Universidade Federal da Para\'{i}ba, Caixa Postal 5008,
58051-970, Jo\~{a}o Pessoa, PB, Brazil}

\author{A. Yu. Petrov}
\email{petrov@fisica.ufpb.br}
\affiliation{Departamento de F\'{i}sica, Universidade Federal da Para\'{i}ba, Caixa Postal 5008,
58051-970, Jo\~{a}o Pessoa, PB, Brazil}

\author{A. P. Ba\^eta Scarpelli}
\email{scarpelli@cefetmg.br}
\affiliation{Centro Federal de Educa\c{c}\~ao Tecnol\'ogica - MG \\
Avenida Amazonas, 7675, 30510-000,  Nova Gameleira - Belo Horizonte, MG, Brazil}

\pacs{11.30.Cp}

%%%%%%%%%
\begin{abstract}
We calculate contributions to the one-loop renormalization in the spinor sector of the minimal Lorentz-violating extended QED in the second order in Lorentz-breaking parameters. From the renormalizability viewpoint, we show that the inclusion of some of the Lorentz-breaking terms in the model is linked to the presence of others. We also demonstrate that the Ward identities are satisfied up to this order.
\end{abstract}
%%%%%%%%%

\maketitle

\section{Introduction}
Nowadays, there is a consensus that the Standard Model of elementary particles is a low energy effective theory for a more fundamental model. The search for this fundamental theory encompasses the study of standard model extensions that show physical meaning and whose low energy limits respect the known experimental results. The Standard Model Extension (SME) \cite{Colladay-Kostelecky}, in its minimal version, is obtained by adding, to the minimal Standard Model, all possible Lorentz-breaking terms that could emerge from spontaneous symmetry breaking at very high energy, which incorporate constant tensors as vacuum expectation values in the process. The upper limit for the magnitude of these background tensors must be fixed by experiments (see \cite{Kost-Russel} for experimental results) and, as a consequence, should be very tiny. The SME is to be understood as an effective description of Lorentz violation at low energy. It is relevant that SME preserves $SU(3)\times SU(2)\times U(1)$ gauge symmetry and renormalizability \cite{Alan2,Colladay3,Colladay,Colladay4}.

In the papers of Samuel and Kosteleck\`y \cite{Samuel1}, \cite{Samuel2} the possibility of Lorentz symmetry violation was first discussed as a natural process when the perturbative string vacuum is unstable. Latter, Carroll, Field and Jackiw presented a first CPT- and Lorentz-violating extension of QED with the inclusion of a Chern-Simons-like term in the photon sector \cite{CFJ}. The Carroll-Field-Jackiw (CFJ) model was exhaustively studied in the subsequent years, mainly concerning its quantum induction from a CPT-odd axial term added to the fermionic part \cite{CFJ-induction}. The so studied CFJ term is a part of the extended QED, which is the subset of the minimal SME that takes care of the Lorentz-violating QED. The extended QED is well established at the tree level and, as mentioned before, was proven to be renormalizable. Besides, a very important subject of study is its quantum dynamics, since the inclusion of these parts in the classical action may cause the radiative induction of new terms. Many papers were dedicated to the investigation of the quantum corrections to the minimal extended QED action with interesting results and discussions, as in the case of the ambiguity of the induced CFJ term. However, these discussions were held almost always up to the first order in the Lorentz-breaking parameters. 

Actually, this focus in the first order correction in the background tensors is justified by the fact that the extended QED corrections to the known results should be very small, so that they can be encompassed by the current experimental error. It is also true, however, that the first order corrections in some of these parameters are null. In these cases, one must pay attention to the lowest-order non-null correction. It is also to be verified the nature of the null corrections: do they vanish just eventually in that order or is there some underlying deeper reason? In \cite{sec-order}, the one-loop corrections to the photon sector of the extended QED at second order in the background tensors were calculated. It was possible to argue that some of the parameters do not induce quantum corrections at all, while others contribute depending on the order of calculation.  It is interesting, for example, to note the cases of the vector $e^\mu$ and the axial vector $f^\mu$, which do not contribute in the first order calculation. The one-loop second-order calculations in $e^\mu$ and $f^\mu$, however, give divergent contributions both to the Maxwell and the CPT-even aether terms. This has relevant implications for the renormalization of extended QED. A first-order one-loop calculation that included only the Lorentz-violating terms with $e^\mu$ and $f^\mu$ might lead us to conclude that the presence in action of an aether-like term is unnecessary. When the second-order results are considered, we see that the aether term must be introduced in the action from the beginning.

This is a compelling observation towards a more complete study of the one-loop renormalization of the extended QED. In the present paper we carry out the renormalization of the fermionic sector of the model up to the second order in Lorentz-violating parameters and investigate the role of the different background tensors in the beta-functions. The paper is divided as follows: in section II, we present the model and discuss the previous results; in section III, we perform the calculation of the fermion self-energy and the vertex correction and check the Ward identities; the beta-functions are calculated in section IV; we discuss our results and present our conclusions in section V.

\section{The model and general discussion on renormalization at one-loop order}

The minimal Lorentz-breaking extended QED, in its most general and renormalizable form, is described by the following classical Lagrangian density  \cite{Alan2}:
\begin{eqnarray}
\label{genrenmod}
{\cal L}&=&\bar{\psi}(i\Gamma^{\nu}D_{\nu}-M)\psi-\frac{1}{4}F_{\mu\nu}F^{\mu\nu}
-\frac{1}{4}\kappa_{\mu\nu\lambda\rho}F^{\mu\nu}F^{\lambda\rho}+%\nonumber\\ &+&
\frac{1}{2}\varepsilon_{\mu\nu\lambda\rho}(k_{AF})^{\mu}A^{\nu}F^{\lambda\rho},
\end{eqnarray}
in which
\begin{equation}
\Gamma^{\nu}=\gamma^{\nu}+c^{\mu\nu}\gamma_{\mu}+d^{\mu\nu}\gamma_{\mu}\gamma_5+e^{\nu}+if^{\nu}\gamma_5+\frac{1}{2}g^{\lambda\mu\nu}\sigma_{\lambda\mu}
\end{equation}
and
\begin{equation}
M=m+im_5 \gamma_5+a_{\mu}\gamma^{\mu}+b_{\mu}\gamma^{\mu}\gamma_5+\frac{1}{2}H^{\mu\nu}\sigma_{\mu\nu}.
\end{equation}
The covariant derivative is given by $D_{\mu}=\partial_{\mu}+iqA_{\mu}$, with $q$ the coupling constant. The  constant tensors (or pseudotensors) $\kappa_{\mu\nu\lambda\rho}$, $(k_{AF})^{\mu}$, $a^{\mu}$, $b^{\mu}$, $c^{\mu\nu}$, $d^{\mu\nu}$, $e^{\mu}$, $f^{\mu}$, $g^{\lambda\mu\nu}$ and $H^{\mu\nu}$ are responsible for the Lorentz-symmetry violation. Concerning $c^{\mu \nu}$ and $d^{\mu \nu}$, here they are treated as symmetric and traceless tensors. The null trace is justified by the fact that the tensors can be redefined so as the traces are absorbed in the Lorentz-invariant part. The simplification of considering $c^{\mu\nu}$ and $d^{\mu\nu}$ symmetric is based on the hypothesis that only these parts of the tensors contribute to physical results. As for the tensor $g^{\mu\nu\alpha}$, it is antisymmetric in the two first indices. Here, for simplicity, we use a particular form of $g^{\mu\nu\lambda}$, given by the completely antisymmetric tensor $g^{\mu\nu\lambda}=\varepsilon^{\mu\nu\lambda\rho}h_\rho$. The $a^\mu$ vector can be eliminated from the action by a suitable redefinition of the fields. All these considerations about the background tensors were discussed in \cite{Colladay-McDonald}, in which it was shown that some parameters can be removed from the Lagrangian by using an appropriate redefinition of the spinor field components. 

In \cite{sec-order}, this model was used to study the one-loop second-order contribution in these parameters to the photon two-point function. We proceed now to the definition of the renormalization and normalization conditions in the model for the purpose of performing an investigation of its second-order one-loop quantum corrections to the fermion two-point function and to the interaction term (three-point function). We write below the Lagrangian for the quantum model, in which the renormalization constants were introduced,
\begin{eqnarray}
\mathcal{L}&=&-\frac{Z_{3}}{4}F_{\mu\nu}F^{\mu\nu} 
-\frac{Z_3}{4}(Z_\kappa)^{\mu\nu\alpha\beta}_{\rho\sigma\lambda\theta}\kappa_{\mu\nu\alpha\beta}F^{\rho\sigma}F^{\lambda\theta}
+Z_{2}\bar{\psi}i\gamma^{\mu}\partial_{\mu}\psi-Z_{2}Z_m m\bar{\psi}\psi-Z_{1}q\bar{\psi}\gamma^{\mu}\psi A_{\mu}\nonumber\\
&+&iZ_2\bar{\psi}\left[\left(Z_{c}\right)_{\,\,\,\,\alpha\beta}^{\nu\mu}c^{\alpha\beta}\gamma_{\nu}+\left(Z_{d}\right)_{\,\,\,\,\alpha\beta}^{\nu\mu}d^{\alpha\beta}\gamma_{5}\gamma_{\nu}+\left(Z_{e}\right)_{\,\,\alpha}^{\nu}e^{\alpha}+i\left(Z_{f}\right)_{\,\,\alpha}^{\mu}f^{\alpha}\gamma_{5}+\frac{1}{2}\left(Z_{g}\right)_{\,\,\,\,\alpha\beta\gamma}^{\lambda\nu\mu}g^{\alpha\beta\gamma}\sigma_{\lambda\nu}\right]\partial_{\mu}\psi\nonumber\\
&-&Z_1q\bar{\psi}\left[\left(Z_{c}\right)_{\,\,\,\,\alpha\beta}^{\nu\mu}c^{\alpha\beta}\gamma_{\nu}
+\left(Z_{d}\right)_{\,\,\,\,\alpha\beta}^{\nu\mu}d^{\alpha\beta}\gamma_{5}\gamma_{\nu}
+\left(Z_{e}\right)_{\,\,\alpha}^{\nu}e^{\alpha}+i\left(Z_{f}\right)_{\,\,\alpha}^{\mu}f^{\alpha}\gamma_{5}
+\frac{1}{2}\left(Z_{g}\right)_{\,\,\,\,\alpha\beta\gamma}^{\lambda\nu\mu}g^{\alpha\beta\gamma}\sigma_{\lambda\nu}\right]\psi A_{\mu},
\label{model}
\end{eqnarray}

Note that the Lorentz-violating (LV) parameters of $M$ are not being considered here. We are restricting our analysis to the parameters contained in $\Gamma^\mu$ and to the $\kappa^{\mu\nu\alpha\beta}$ tensor of the CPT-even term of the gauge sector.  The latter is necessary, as it receives divergent contributions from several of the $\Gamma^\mu$ parameters, as shown in \cite{sec-order}. Besides, it contributes to the beta-function of the parameter $c^{\mu\nu}$ already at first order \cite{Alan2}. Note also that, in principle, we do not assume any relationship between the renormalization constants from the Ward identities, such as f.e. $Z_1 = Z_2$. We have left for the next section the explicit verification of this identity at one-loop and up to second-order in the Lorentz-violating parameters. We define the counterterm Lagrangian by
\begin{eqnarray}
\mathcal{L}_{ct}&=&-\frac{\delta_{3}}{4}F_{\mu\nu}F^{\mu\nu}
-\frac{1}{4}(\delta_\kappa)^{\mu\nu\alpha\beta}_{\rho\sigma\lambda\theta}\kappa_{\mu\nu\alpha\beta}F^{\rho\sigma}F^{\lambda\theta}
+\delta_{2}i\gamma^{\mu}\bar{\psi}\partial_{\mu}\psi-\delta_{m}m\bar{\psi}\psi
-\delta_{1}q\left(\bar{\psi}\gamma^{\mu}\psi\right)A_{\mu}\nonumber\\
&+&i\bar{\psi}\left[\left(\delta_{c}\right)_{\alpha\beta}^{\nu\mu}c^{\alpha\beta}\gamma_{\nu}
+\left(\delta_{d}\right)_{\alpha\beta}^{\nu\mu}d^{\alpha\beta}\gamma_{5}\gamma_{\nu}
+\left(\delta_{e}\right)_{\alpha}^{\mu}e^{\alpha}+i\left(\delta_{f}\right)_{\alpha}^{\mu}f^{\alpha}\gamma_{5}
+\frac{1}{2}\left(\delta_{g}\right)_{\alpha\beta\gamma}^{\lambda\nu\mu}g^{\alpha\beta\gamma}\sigma_{\lambda\nu}\right]
\partial_{\mu}\psi\nonumber\\
&-&q\bar{\psi}\left[\left(\overline{\delta}_{c}\right)_{\alpha\beta}^{\nu\mu}c^{\alpha\beta}\gamma_{\nu}
+\left(\overline{\delta}_{d}\right)_{\alpha\beta}^{\nu\mu}d^{\alpha\beta}\gamma_{5}\gamma_{\nu}
+\left(\overline{\delta}_{e}\right)_{\alpha}^{\mu}e^{\alpha}+i\left(\overline{\delta}_{f}\right)_{\alpha}^{\mu}f^{\alpha}\gamma_{5}
+\frac{1}{2}\left(\overline{\delta}_{g}\right)_{\alpha\beta\gamma}^{\lambda\nu\mu}g^{\alpha\beta\gamma}\sigma_{\lambda\nu}\right]\psi A_{\mu},
\label{counterterms}
\end{eqnarray}
where
\begin{equation}
\delta_{1}=Z_{1}-1,\,\,\,\,\,\,\,\,\,\, \delta_{2}=Z_{2}-1,\,\,\,\,\,\,\,\,\,\,\,\delta_{3}=Z_{3}-1,\,\,\,\,\,\,\,\,\,\delta_{m}=Z_m-1,\nonumber
\end{equation}

\begin{equation}
\left(\delta_{c}\right)_{\alpha\beta}^{\nu\mu}=Z_2\left(Z_{c}\right)_{\,\,\,\,\alpha\beta}^{\nu\mu}
-\delta_{\alpha}^{\nu}\delta_{\beta}^{\mu},\,\,\,\,\,\,\,\,\,\,
\left(\delta_{d}\right)_{\alpha\beta}^{\nu\mu}=Z_2\left(Z_{d}\right)_{\,\,\,\,\alpha\beta}^{\nu\mu}
-\delta_{\alpha}^{\nu}\delta_{\beta}^{\mu},\,\,\,\,\,\,\,\,\,\,
\left(\delta_{e}\right)_{\alpha}^{\mu}=Z_2\left(Z_{e}\right)_{\,\,\alpha}^{\nu}-\delta_{\alpha}^{\nu},
\nonumber
\end{equation}

\begin{equation}
\left(\delta_{f}\right)_{\alpha}^{\mu}=Z_2\left(Z_{f}\right)_{\,\,\alpha}^{\nu}-\delta_{\alpha}^{\nu}, \,\,\,\,\,\,\,\,\,\, \left(\delta_{g}\right)_{\alpha\beta\gamma}^{\lambda\nu\mu}=Z_2\left(Z_{g}\right)_{\alpha\beta\gamma}^{\lambda\nu\mu}-\delta_{\alpha}^{\lambda}\delta_{\beta}^{\nu}\delta_{\gamma}^{\mu},\nonumber
\end{equation}

\begin{equation}
\left(\overline{\delta}_{c}\right)_{\alpha\beta}^{\nu\mu}=Z_1\left(Z_{c}\right)_{\,\,\,\,\alpha\beta}^{\nu\mu}
-\delta_{\alpha}^{\nu}\delta_{\beta}^{\mu},\,\,\,\,\,\,\,\,\,\,
\left(\overline{\delta}_{d}\right)_{\alpha\beta}^{\nu\mu}=Z_1\left(Z_{d}\right)_{\,\,\,\,\alpha\beta}^{\nu\mu}
-\delta_{\alpha}^{\nu}\delta_{\beta}^{\mu}, \,\,\,\,\,\,\,\,\,\,\,\,
\left(\overline{\delta}_{e}\right)_{\alpha}^{\mu}=Z_1\left(Z_{e}\right)_{\,\,\alpha}^{\nu}-\delta_{\alpha}^{\nu},
\nonumber
\end{equation}

\begin{equation}
\left(\overline{\delta}_{f}\right)_{\alpha}^{\mu}=Z_1\left(Z_{f}\right)_{\,\,\alpha}^{\nu}-\delta_{\alpha}^{\nu}
\,\,\,\,\, \mbox{and} \,\,\,\,\, \left(\overline{\delta}_{g}\right)_{\alpha\beta\gamma}^{\lambda\nu\mu}=
Z_1\left(Z_{g}\right)_{\alpha\beta\gamma}^{\lambda\nu\mu}-\delta_{\alpha}^{\lambda}\delta_{\beta}^{\nu}\delta_{\gamma}^{\mu}.\nonumber
\end{equation}
The Ward identity, if respected, will set $Z_1=Z_2$ and $(\delta_x)^J=(\overline{\delta}_x)^J$, in which $x$ represents the Lorentz-breaking parameter and $J$ indicates the appropriated set of Lorentz indices. 

We write below in this section the general expressions for the fermion two-point function and vertex correction and let for subsequent sections the explicit results. The renormalized two-point function of the fermion field can be written as
\begin{equation}
i\Sigma^R(p)=i\Sigma(p)+i\Sigma_{ct}(p),
\label{sigmaloop}
\end{equation}
in which $i\Sigma(p)$ has the general form
\begin{equation}
i\Sigma(p)=I_{log}(\lambda^{2})\left\{A_{\psi}\cancel{p} + A_{m}m + A_{e}^{\mu}p_{\mu}+i A_f^\mu \gamma_5 p_\mu +  A_{c}^{\mu\nu}\gamma_{\mu}p_{\nu} + A_{d}^{\mu\nu}\gamma_5\gamma_{\mu}p_{\nu} + \frac 12 A_{g}^{\alpha\beta\mu}\sigma_{\alpha\beta}p_{\mu}\right\}
\label{twoPoint}
\end{equation}
and $i\Sigma_{ct}(p)$ is the contribution from the counterterms. The coefficients $A_{x}^J$, at one-loop order, are given by
\begin{equation}
A_x^J=\left\{(\rho_0)^J_x+(\rho_1)^J_x+(\rho_2)^J_x+\cdots\right\}q^2,
\end{equation}
where $(\rho_i)^J_x$ is the term of $A_x^J$ of $i$-th order in the parameters $(x_1^{J_1}, x_2^{J_2}\cdots)$.
Here we are using the notation of Implicit Regularization \cite{IR},
\begin{equation}
I_{log}(\lambda^{2}) = \int^\Lambda \frac{d^{4}k}{(2\pi)^4}\frac{1}{(k^2-\lambda^2)^{2}},
\end{equation}
for the logarithmic divergence, in which we assume the presence of a regularization, indicated by the superscript $\Lambda$, and $\lambda^2$ is a mass parameter which can be introduced by means of the scale relation
\begin{equation}
I_{log}(\lambda^2)=I_{log}(m^2)-\frac{i}{16\pi^2}\ln{\left(\frac{\lambda^2}{m^2}\right)}.
\end{equation}

Up to second order in the parameters  $x_i^J$, we have
\begin{equation}
i\Sigma(p)=i\Sigma^{(0)}(p)+i\Sigma^{(1)}(p)+i\Sigma^{(2)}(p),
\end{equation}
whose contributions can be extracted from the expansion up to second order in the parameters of a unique graphic with vertices $-iq\Gamma^\mu$ and the modified propagators for the fermion and the photon. Alternatively, the terms with the background tensors can be treated as vertices which are inserted in the graphs. For example, the second-order fermion two-point function is obtained after the summation of the Feynman diagrams of Fig. \ref{e_parameter}. The counterterms that cancel the logarithmic divergences in (\ref{twoPoint}) are contained in $i\Sigma_{ct}(p)$, which is built from (\ref{counterterms}) and reads
\begin{eqnarray}
i\Sigma_{ct}(p)&=&i\delta_{2}\cancel{p}-i\delta_{m} + i\left(\delta_{e}\right)_{\alpha}^{\mu}e^{\alpha}p_{\mu} - \left(\delta_{f}\right)_{\alpha}^{\mu}f^{\alpha}\gamma_{5}p_{\mu} + i\left(\delta_{c}\right)_{\alpha\beta}^{\nu\mu}
c^{\alpha\beta}\gamma_{\mu}p_{\nu} +i\left(\delta_{d}\right)_{\alpha\beta}^{\nu\mu}d^{\alpha\beta}\gamma_{5}\gamma_{\nu}p_{\mu} \nonumber \\
&+& \frac{i}{2}\left(\delta_{g}\right)_{\alpha\beta\gamma}^{\lambda\nu\mu}g^{\alpha\beta\gamma}\sigma_{\lambda\nu}
\label{twoPointCount}
\end{eqnarray}

The renormalized vertex correction at the one-loop order is written as
\begin{equation}
-iq\Lambda_R^{\mu}=-iq\Lambda^{\mu}-iq\Lambda_{ct}^{\mu}.
\end{equation}
The first term, $-iq\Lambda^{\mu}$, on the right-hand-side of the equation above, has the general form
\begin{equation}
-iq\Lambda^{\mu}=I_{log}(m^{2})\left\{B_{\psi}\gamma^{\mu}+ B_{e}^{\mu}+iB_{f}^{\mu}\gamma_{5}+B_{c}^{\nu\mu}\gamma_{\nu}+ B_{d}^{\nu\mu}\gamma_{5}\gamma_{\nu} + \frac 12
B_{g}^{\lambda\nu\mu}\sigma_{\lambda\nu}\right\}
\label{threePoint}
\end{equation}
and, from (\ref{counterterms}), the amplitude for the counterterms, $-iq\Lambda_{ct}^{\mu}$, is  given by
\begin{equation}
-iq\Lambda_{ct}^{\mu}=-iq\left(\overline{\delta}_{q}\gamma^{\mu} + \left(\overline{\delta}_{e}\right)_{\alpha}^{\mu}e^{\alpha} + i\left(\overline{\delta}_{f}\right)_{\alpha}^{\mu}f^{\alpha}\gamma_{5} + \left(\overline{\delta}_{c}\right)_{\alpha\beta}^{\nu\mu}c^{\alpha\beta}
\gamma_{\nu}+\left(\overline{\delta}_{d}\right)_{\alpha\beta}^{\nu\mu}d^{\alpha\beta}\gamma_{5}\gamma_{\nu} + \frac{1}{2}\left(\overline{\delta}_{g}\right)_{\alpha\beta\gamma}^{\lambda\nu\mu}g^{\alpha\beta\gamma}\sigma_{\lambda\nu}\right).
\label{threePointCount}
\end{equation}

The coefficients $B_{x}^J$, at one-loop order, are given by
\begin{equation}
B_x^J=\left\{(\sigma_0)^J_x+(\sigma_1)^J_x+(\sigma_2)^J_x+\cdots\right\}q^3,
\end{equation}
in which $(\sigma_i)^J_x$ is the term of $B_x^J$ of $i$-th order in the parameters $(x_1^{J_1}, x_2^{J_2}\cdots)$. Up to second order in the Lorentz-breaking parameters, we have
\begin{equation}
-i q\Lambda_\mu(p,p')=-iq\Lambda_\mu^{(0)}(p,p')-iq\Lambda_\mu^{(1)}(p,p')-iq\Lambda_\mu^{(2)}(p,p').
\end{equation}
For example, the second-order contribution $-iq\Lambda_\mu^{(2)}$ is obtained by the calculation of the Feynman diagrams shown in Fig. \ref{e_parameter_three_Point}. The coefficients $A_x^J$ and $B_x^J$ must be such that the Ward identity
\begin{equation}
(p-p')_\lambda \Lambda^\lambda_{1loop}=\Sigma_{1loop}(p)-\Sigma_{1loop}(p')
\label{Ward identity}
\end{equation}
is respected ($l=p-p'$ is the momentum of the outgoing photon). In particular, from (\ref{twoPoint}), (\ref{twoPointCount}), (\ref{threePoint}) and (\ref{threePointCount}), we can see that the renormalization constants $Z_{1}$ and $Z_{2}$, if we adopt a subtraction scheme that cancels only the $I_{log}(\lambda^2)$'s, are given by
\begin{eqnarray}
Z_{1} = 1 - iB_\psi q^{-1}I_{log}(\lambda^2)\,\,\,\,\,\text{and}\,\,\,\,\, Z_{2} = 1+i A_\psi I_{log}(\lambda^2).
\end{eqnarray}
As we shall see in the next section, the explicit calculation of the diagrams will give us the expected result
\begin{equation}
A_{\psi} = -q^{-1}B_{\psi},
\label{identCoeff}
\end{equation}
so that we have that, up to the second-order in the parameters, the identity $Z_{1} = Z_{2}$ still holds in the minimal Lorentz-violating extension of QED. We also obtain, for the other renormalization constants,
\begin{equation}
Z_2(Z_x)^J_{J'}x^{J'}=x^J+iA_x^J I_{log}(\lambda^2)
\end{equation}
and
\begin{equation}
Z_1(Z_x)^J_{J'}x^{J'}=x^J-iq^{-1}B_x^J I_{log}(\lambda^2).
\end{equation}
From the equations above, we have the conditions
\begin{eqnarray}
A_x^J=-q^{-1}B_x^J.
\end{eqnarray}

In the next section, we present the explicit calculations of the divergent parts of the fermion self-energy and the vertex correction, as well as we exhibit the results for the vacuum polarization tensor.

\section{Two- and three-point  functions}
In this section we will present explicit results for the amplitudes given by the summation of the diagrams in Figs. \ref{e_parameter} and \ref{e_parameter_three_Point} (two- and three-point functions, respectively). The gauge field propagator in the Feynman gauge, represented by the wavy lines, and the fermion propagator, drawn with solid lines in the diagrams, are given, respectively, by
\begin{equation}
\Delta_0^{\mu\nu}(p) = -\frac{i\eta^{\mu\nu}}{p^{2}}
\label{gauge}
\end{equation}
and
\begin{equation}
S_{0}(p)=\frac{i}{\cancel{p} - m},  
\end{equation}
in which the subscript ``$0$'' is to indicate the zeroth-order contribution in the Lorentz-violating parameters and the adopted metric is $\eta_{\mu\nu} = (1,-1,-1,-1)$. Since we are treating perturbatively the background tensors, they will appear as vertices in the Feynman diagrams. So, a cross appearing in the diagrams stands for
\begin{equation}
ip_{\mu}\Gamma^\mu_1
\label{ruleSelf}
\end{equation}
in the bilinear $\bar{\psi}\psi$ vertex,
\begin{equation}
-iq\Gamma^\mu_1
\label{ruleVertex}
\end{equation}
in the trilinear vertex $\bar{\psi}\psi A^{\mu}$ and
\begin{equation}
-2ip_\alpha p_\beta \kappa^{\alpha\mu\beta\nu}
\label{ruleAA}
\end{equation}
in the bilinear $A^\mu A^\nu$ vertex. Besides,
\begin{equation}
\Gamma^\nu_1=c^{\mu\nu}\gamma_{\mu} + d^{\mu\nu}\gamma_{\mu}\gamma_5 + e^{\nu} + if^{\nu}\gamma_5 + \frac{1}{2}g^{\lambda\mu\nu}\sigma_{\lambda\mu}
\end{equation}
is the Lorentz-violating part of $\Gamma^\mu$.

With the help of the Feynman rules (\ref{gauge})-(\ref{ruleAA}), we calculate all the divergent diagrams which contribute at the second order in the parameters $c^{\mu\nu}$, $d^{\mu\nu}$, $e^{\mu}$, $f^{\mu}$ and $g^{\mu\nu\rho}$ to the fermion two-point function and to the vertex correction in the extended QED. 

\subsection {Self-energy of $\psi$} 
Let us first consider the corrections at second order in the parameters. We have contributions with insertions of $\Gamma_1$ both in the fermion line and in the vertex $\bar{\psi} A_\mu \psi$, whereas the tensor $\kappa^{\mu\nu\alpha\beta}$ in inserted only in the photon propagator. The diagrams which contribute are displayed in Fig. \ref{e_parameter}, and the corresponding amplitudes are given below:

\begin{figure}[h]
	\begin{center}
		\includegraphics[scale=0.5]{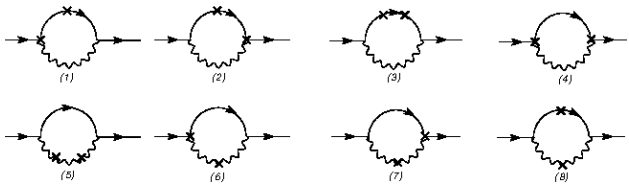}
		\caption{Diagrammatic representation of the two-point function at second-order in the Lorentz-violating parameters. The wavy and solid lines represent the photon and fermion propagators and the crosses indicate the insertions.}
		\label{e_parameter}
	\end{center}
\end{figure}

\begin{eqnarray}
i\Sigma^{(2)}_1(p)&=&q^2\int\frac{d^4l}{(2\pi)^4}\,
\frac{(p-l)_\alpha\gamma^\mu(\cancel{p}-\cancel{l}+m)\Gamma_1^\alpha (\cancel{p}-\cancel{l}+m) \Gamma_{1\mu}}
{l^2\left[(p-l)^2-m^2\right]^{2}};\nonumber\\
i\Sigma^{(2)}_2(p)&=&q^2\int\frac{d^4l}{(2\pi)^4}\,
\frac{(p-l)_\alpha\Gamma_1^\mu(\cancel{p}-\cancel{l}+m)\Gamma_1^\alpha (\cancel{p}-\cancel{l}+m) \gamma_\mu}
{l^2\left[(p-l)^2-m^2\right]^{2}};\nonumber\\
i\Sigma^{(2)}_3(p)&=&-q^2\int\frac{d^4l}{(2\pi)^4}\,
\frac{(p-l)_\alpha(p-l)_\beta\gamma^\mu(\cancel{p}-\cancel{l}+m)\Gamma_1^\alpha (\cancel{p}-\cancel{l}+m)\Gamma_1^\beta
(\cancel{p}-\cancel{l}+m)\gamma_{\mu}}
{l^2\left[(p-l)^2-m^2\right]^{3}};\nonumber\\
i\Sigma^{(2)}_4(p)&=&-q^2\int\frac{d^4l}{(2\pi)^4}\,
\frac{\Gamma_1^\mu(\cancel{p}-\cancel{l}+m)\Gamma_{1\mu}}
{l^2\left[(p-l)^2-m^2\right]}, \nonumber \\
i\Sigma^{(2)}_5(p)&=&-4q^2  \kappa^{\lambda\alpha\delta\beta}
\kappa^{\tau\sigma\theta}_{\,\,\,\,\,\,\,\,\,\beta}
\int\frac{d^4l}{(2\pi)^4}\,
\frac{\gamma_\alpha(\cancel{p}-\cancel{l}+m)\gamma_\sigma l_\lambda l_\delta l_\theta l_\tau}
{l^6\left[(p-l)^2-m^2\right]},\nonumber \\
i\Sigma^{(2)}_6(p)&=&2q^2  \kappa^{\mu\alpha\nu\beta}\int\frac{d^4l}{(2\pi)^4}\,
\frac{\gamma_\nu(\cancel{p}-\cancel{l}+m)\Gamma_{1\mu} l_\alpha l_\beta}
{l^4\left[(p-l)^2-m^2\right]}, \nonumber \\
i\Sigma^{(2)}_7(p)&=&2q^2  \kappa^{\mu\alpha\nu\beta}\int\frac{d^4l}{(2\pi)^4}\,
\frac{\Gamma_{1\nu}(\cancel{p}-\cancel{l}+m)\gamma_\mu l_\alpha l_\beta}
{l^4\left[(p-l)^2-m^2\right]},\nonumber \\
i\Sigma^{(2)}_8(p)&=&-2q^2  \kappa^{\mu\alpha\nu\beta}\int\frac{d^4l}{(2\pi)^4}\,
\frac{(p-k)_\rho\gamma_{\nu}(\cancel{p}-\cancel{l}+m)\Gamma_{1}^\rho(\cancel{p}-\cancel{l}+m)\gamma_{\mu}l_\alpha l_\beta}
{l^4\left[(p-l)^2-m^2\right]^2}.
\end{eqnarray}
After the calculation of the Feynman integrals, we obtain the divergent second-order contributions as
\begin{equation}
i\Sigma^{(2e)}(p) = -\frac{1}{3}q^{2}I_{log}(\lambda^{2})\Big\{e^2 \left(\cancel{p}+9m\right) - 4 \cancel{e} (e \cdot p)\Big\} ;
\label{Se}
\end{equation}
\begin{equation}
i\Sigma^{(2f)}(p)= -\frac{1}{3}q^{2}I_{log}(\lambda^{2})\Big\{f^{2}\cancel{p} - 4\cancel{f}(f \cdot p)\Big\};    
\end{equation}

\begin{equation}
i\Sigma^{(2h)}(p)= -\frac{1}{3}q^2 I_{log}(\lambda^2)\Big\{h^2\left(\cancel{p}+9m\right)-4\cancel{h} (h \cdot p)\Big\};    
\end{equation}

\begin{eqnarray}
i\Sigma^{(2c)}(p)= \frac{1}{6}q^{2}I_{log}(\lambda^{2})\Big\{-8m\,c_{\mu\nu}c^{\mu\nu}+3 c_{\mu\nu}c^{\mu\nu}\cancel{p}
-12c_{\mu\sigma}c_{\nu}^\sigma p^{\mu}\gamma^{\nu}\Big\};
\end{eqnarray}

\begin{eqnarray}
i\Sigma^{(2d)}(p)= \frac{1}{6}q^{2}I_{log}(\lambda^{2})\Big\{16m\, d_{\mu\nu}d^{\mu\nu}+3d_{\mu\nu}d^{\mu\nu}\cancel{p}
-12d_{\mu\sigma}d_{\nu}^\sigma p^{\mu}\gamma^{\nu}\Big\};
\label{Sd}
\end{eqnarray}
\begin{eqnarray}
&&i\Sigma^{(2\kappa)}(p)=-\frac{1}{24} q^2 I_{log}(\lambda^2) \kappa^{\lambda\alpha\delta\beta}
\kappa^{\tau\sigma\theta}_{\,\,\,\,\,\,\,\,\,\beta}
\Big\{\left[(-3\cancel{p}+4m)\eta_{\alpha\sigma} + 6 \gamma_\alpha p_\sigma \right]\eta_{(\lambda\delta}\eta_{\theta\tau)}
\nonumber \\
&& + \eta_{\alpha\sigma} p_{(\lambda}\gamma_\delta\eta_{\theta\tau)}
\Big\},
\label{kappa2}
\end{eqnarray}
in which the indices limited by parentheses must be exchanged symmetrically without the factor $1/n!$. 
With the results above, discarding for awhile the crossed terms, we obtain the total divergent part of the second-order correction to the fermion self-energy:
\begin{equation}
i\Sigma^{(2)}(p) = i\Sigma^{(2c)}(p) + i\Sigma^{(2d)}(p) + i\Sigma^{(2e)} +i\Sigma^{(2f)}(p) + i\Sigma^{(2g)}(p)
+ i\Sigma^{(2\kappa)}(p).
\label{twopointsum}
\end{equation}
The zeroth and first-order divergent corrections are given by
\begin{equation}
i\Sigma^{(0)}(p)=q^2 I_{log}(\lambda^2)\left(\cancel{p}-4m\right)
\end{equation}
and
\begin{eqnarray}
&&i\Sigma^{(1)}(p)=\frac{1}{6}q^2 I_{log}(\lambda^2)\Big\{6 e^\mu \left(p_\mu - 3m \gamma_\mu \right) + 6i(f\cdot p)\gamma_5 
- 10 c^{\nu\mu}\gamma_\nu p_\mu - 10 d^{\nu\mu}\gamma_\nu \gamma_5 p_\mu \\ \nonumber
&& + 3i \varepsilon^{\lambda\beta\rho\alpha}h_\alpha \gamma_\lambda\gamma_\beta \left(p_\rho-m \gamma_\rho\right)
+ 8 \gamma_\alpha p_\beta \kappa_\rho\,^{\beta\rho\alpha}\Big\}.
\end{eqnarray}

For example, the coefficient $A_\psi$, which is needed for obtaining the renormalization constant $Z_2$ is given by
\begin{equation}
A_{\psi}=q^{2}\left\{1-\frac{1}{24}\left(8e^{2} + 8f^{2} + 8h^{2} - 12c_{\mu\nu}c^{\mu\nu} - 12 d_{\mu\nu}d^{\mu\nu} - 3\eta_{(\lambda\delta} \eta_{\theta\tau)}\kappa^{\beta\delta\alpha\lambda}\kappa_{\beta \,\, \alpha}^{\,\,\,\,\theta \,\,\tau}\right)\right\}.
\label{apsi}
\end{equation}
The other coefficients are found to be
\begin{equation}
A_m=-q^2\left\{4 +\frac{1}{2}\left(6\cancel{e}+i\varepsilon^{\lambda\beta\rho\alpha}h_\alpha \gamma_\lambda 
\gamma_\beta \gamma_\rho \right) + \frac{1}{6}\left(18 e^2 + 18h^2 +8 c_{\mu\nu}c^{\mu\nu}-16d_{\mu\nu}d^{\mu\nu}
+\eta_{(\lambda\delta} \eta_{\theta\tau)}\kappa^{\beta\delta\alpha\lambda}\kappa_{\beta \,\, \alpha}^{\,\,\,\,\theta \,\,\tau}\right)\right\},
\label{am}
\end{equation}
\begin{eqnarray}
&&A_c^{\mu\nu}=\frac{1}{3}q^2\Big\{\left(-5c^{\mu\nu} + 4 \kappa^{\mu\theta\nu}_{\,\,\,\,\,\,\,\,\theta}\right)+ \Big\{4e^\mu e^\nu + 4f^\mu f^\nu + 4 h^\mu h^\nu - 6 c^{\mu\sigma}c^{\nu}_\sigma 
- 6 d^{\mu\sigma}d^{\nu}_\sigma + G^{\mu\nu}(\kappa)\Big\} \Big\},
\label{ac}
\end{eqnarray}
\begin{eqnarray}
A^\mu_e=q^2 e^\mu, \,\,\,\, A^\mu_f=q^2 f^\mu, \,\,\,\, A^{\mu\nu}_d=-\frac 53 q^2 d^{\mu\nu} \,\,\,\,
\mbox{and} \,\,\,\, A^{\mu\nu\alpha}_g=q^2 g^{\mu\nu\alpha},
\label{as}
\end{eqnarray}
with
\begin{equation}
G^{\mu\nu}(\kappa)=-\frac 14 \left[3\kappa^{\mu\lambda\delta\beta}
\kappa^{\nu\tau\theta}_{\,\,\,\,\,\,\,\,\,\beta}  \eta_{(\lambda\delta}\eta_{\theta\tau)}
+ 2\kappa^{\mu\alpha\nu\beta}\kappa^\theta_{\,\,\alpha\theta\beta} 
+ 4 \kappa^{\mu\alpha\theta\beta}\kappa^\nu_{\,\,\alpha\theta\beta} \right].
\end{equation}
Note that the coefficients $A_e^\mu$, $A_f^\mu$, $A_d^{\mu\nu}$ and $A_g^{\mu\nu\alpha}$ do not have second-order contributions.

\subsection{Three-point vertex function $\bar{\psi}\psi A^{\mu}$}
\begin{figure}[h]
	\begin{center}
		\includegraphics[scale=0.8]{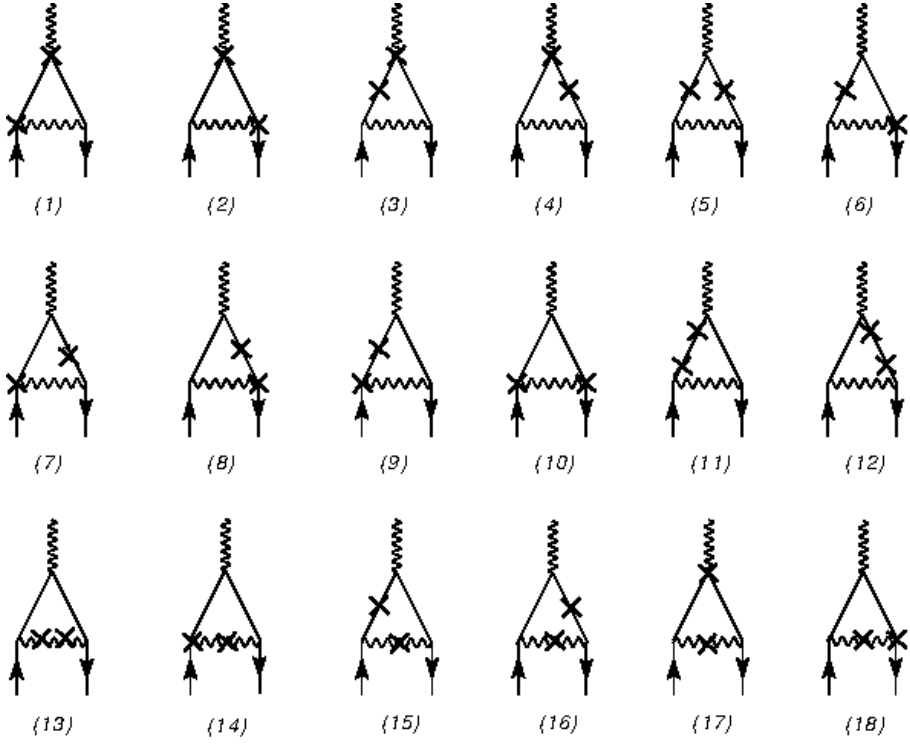}
		\caption{Diagrammatic representation of the one-loop three-point function at second-order in the parameters. The wavy and solid lines represent the photon and fermion propagators, respectively, and the crosses indicate the insertions of the background tensors.}
		\label{e_parameter_three_Point}
	\end{center}
\end{figure}

The one-loop Feynman diagrams that contribute to the vertex function at second-order in the  Lorentz-violating parameters are shown in Fig. \ref{e_parameter_three_Point}. As for the fermion self-energy amplitudes presented before, the background tensors are considered as vertices inserted in the diagrams, represented by crosses. The sum of the diagrams which contribute at second-order in each one of the parameters reads

\begin{eqnarray}
-iq\Lambda^{(2e)\lambda}&=& \frac 13 q^3I_{log}(\lambda^{2})\left\{\gamma^{\lambda}e^{2} - 4 e^{\lambda} \cancel{e}\right\},\label{graphe}\\   
-iq\Lambda^{(2f)\lambda}&=& \frac 13 q^3 I_{log}(\lambda^{2})\left\{\gamma^{\lambda}f^{2} - 4 f^{\lambda} \cancel{f}\right\},\\ 
-iq\Lambda^{(2h)\lambda}&=& \frac 13 q^3 I_{log}(\lambda^{2})\left\{\gamma^{\lambda}h^{2} - 4 h^{\lambda} \cancel{h}\right\}, \label{graphg} \\
-iq\Lambda^{(2c)\lambda}&=& - \frac 12 q^{3}I_{log}(\lambda^{2})\left\{\gamma^{\lambda}c_{\mu\nu}c^{\mu\nu} - 4\gamma^{\sigma}c^{\mu\lambda}c_{\mu\sigma}\right\}\label{graphc}, \\
-iq\Lambda^{(2d)\lambda}&=& - \frac 12 q^3 I_{log}(\lambda^2)\left\{\gamma^{\lambda}d_{\mu\nu}d^{\mu\nu} - 4\gamma^{\sigma}d^{\mu\lambda}d_{\mu\sigma}\right\} \,\,\,\,\,\, \mbox{and} \\
-iq\Lambda^{(2\kappa)\lambda}&=& \frac{1}{24}q^3 I_{log}(\lambda^{2})
\kappa^{\gamma\sigma\tau\delta}\kappa^{\theta\nu\lambda}_{\,\,\,\,\,\,\,\,\,\delta} 
\left\{\eta_{\sigma\nu}\delta^\mu_{(\lambda}\gamma_{\theta}\eta_{\tau\sigma)}+3\left(2\gamma_\sigma\delta^\mu_\nu -\gamma^\mu\eta_{\sigma\nu}\right)\eta_{(\theta\lambda}\eta_{\tau\gamma)}\right\},
\end{eqnarray}
while the total second-order one-loop vertex correction, excluding the crossed terms, is written as the sum
\begin{equation}
-iq\Lambda_\mu^{(2)} = -iq\left(\Lambda_\mu^{(2c)} + \Lambda_\mu^{(2d)} +\Lambda_\mu^{(2e)} + \Lambda_\mu^{(2f)} + \Lambda_\mu^{(2g)} + \Lambda_\mu^{(2\kappa)}\right).
\label{threepointsum}
\end{equation}
We also need the zero-th and first order contributions, given by
\begin{equation}
-iq\Lambda_\mu^{(0)}=-q^3 I_{log}(\lambda^2)\gamma_\mu
\end{equation}
and
\begin{equation}
-iq\Lambda_\mu^{(1)}=\frac 16 q^3 I_{log}(\lambda^2)\left\{- 6 e_\mu - 6 i f_\mu \gamma_5 + 10 c_{\rho\mu}\gamma^\rho + 
10 d_{\rho\mu}\gamma^\rho \gamma_5 - 3i \varepsilon_{\sigma\beta\mu\alpha}h^\alpha \gamma^\sigma \gamma^\beta 
-8\kappa^{\theta\mu\,\,\,\,\sigma}_{\,\,\,\,\,\,\theta} \gamma_\sigma\right\}.
\end{equation}
From the above results, we obtain
\begin{equation}
B_{\psi}=-q^{3}\left\{1-\frac{1}{24}\left(8e^{2} + 8f^{2} + 8h^{2} - 12c_{\mu\nu}c^{\mu\nu} - 12 d_{\mu\nu}d^{\mu\nu} - 3\eta_{(\lambda\delta} \eta_{\theta\tau)}\kappa^{\beta\delta\alpha\lambda}\kappa_{\beta \,\, \alpha}^{\,\,\,\,\theta \,\,\tau}\right)\right\}  
\end{equation}
and
\begin{equation}
B_c^{\mu\nu}=-\frac{1}{3}q^3\left\{\left(-5c^{\mu\nu} + 4 \kappa^{\mu\theta\nu}_{\,\,\,\,\,\,\,\,\theta}\right) + \Big\{4e^\mu e^\nu + 4f^\mu f^\nu + 4 h^\mu h^\nu - 6 c^{\mu\sigma}c^{\nu}_\sigma 
- 6 d^{\mu\sigma}d^{\nu}_\sigma + G^{\mu\nu}(\kappa)\Big\} \right\}.
\end{equation}
The other coefficients are just of first-order:
\begin{eqnarray}
B^\mu_e=-q^3 e^\mu, \,\,\,\, B^\mu_f=-q^3 f^\mu, \,\,\,\, B^{\mu\nu}_d=\frac 53 q^3 d^{\mu\nu} \,\,\,\,
\mbox{and} \,\,\,\, B^{\mu\nu\alpha}_g=-q^3 g^{\mu\nu\alpha}.
\end{eqnarray}
All the coefficients are such that $B^J_x=-q A^J_x$, as argued in the last section as the conditions for having the Ward identity (\ref{Ward identity}) satisfied.

\subsection{General results}
\label{general}
There is another approach to the renormalization of extended QED that is more general and compact. Let us write the fermionic part of the quantum Lagrangian density in the form
\begin{equation}
{\cal L}_\psi=iZ_2 \bar{\psi}(Z_\Gamma)^\mu_\nu \Gamma^\nu \partial_\mu \psi -Z_1 q \bar{\psi}(Z_\Gamma)^\mu_\nu \Gamma^\nu \psi A_\mu 
-Z_2 Z_m m \bar{\psi}\psi
\end{equation} 
with the corresponding counterterm Lagrangian given by
\begin{equation}
{\cal L}_{\psi ct}=i \bar{\psi}(\delta_\Gamma)^\mu_\nu \Gamma^\nu \partial_\mu \psi - q \bar{\psi}(\bar{\delta}_\Gamma)^\mu_\nu \Gamma^\nu \psi A_\mu 
-\delta_m m \bar{\psi}\psi,
\end{equation} 
in which
\begin{eqnarray}
(\delta_\Gamma)^\mu_\nu=Z_2 (Z_\Gamma)^\mu_\nu - \delta^\mu_\nu \,\,\,\,\,\, \mbox{and} \,\,\,\,\,\, 
(\bar{\delta}_\Gamma)^\mu_\nu=Z_1 (Z_\Gamma)^\mu_\nu - \delta^\mu_\nu.
\end{eqnarray}
The fermion self-energy and the three-point function will then be written as
\begin{equation}
i\Sigma(p)=I_{log}(\lambda^2)\left(A_\Gamma^\mu p_\mu + A_m m \right)
\end{equation}
and
\begin{equation}
-iq \Lambda^\mu=I_{log}(\lambda^2)B_\Gamma^\mu.
\end{equation}
It is possible to carry out the calculations of the divergent part of the graphs without explicit form of $\Gamma_1^\mu$. For the four first and the last three graphs of Figure 1, one obtains, respectively,
\begin{eqnarray}
&&i\Sigma^{(2)}_{1-4}(p)=\frac{q^2}{96}I_{log}(\lambda^2)\Big[8p_{(\mu}\eta_{\nu\rho)}\Big(\Gamma_1^\alpha \gamma^\rho 
\{\Gamma_1^\mu,\gamma^\nu\}\gamma_\alpha + \gamma_\alpha \gamma^\rho \{\Gamma_1^\mu,\gamma^\nu\} \Gamma_1^\alpha 
+ \gamma^\alpha \Gamma_1^\rho \{\Gamma_1^\mu,\gamma^\nu\} \gamma_\alpha 
+ \gamma^\alpha \gamma^\rho \Gamma_1^\mu \Gamma_1^\nu \gamma_\alpha \Big) + \nonumber \\
&& - p_{(\mu}\eta_{\nu\rho}\eta_{\sigma\theta)}\gamma^\alpha \gamma^\theta \{\Gamma_1^\mu,\gamma^\nu\}\{\Gamma_1^\rho,\gamma^\sigma\}\gamma_\alpha 
- 48 p_\mu\left(\Gamma_1^\alpha\gamma^\mu\Gamma_{1\alpha} + \Gamma_1^\alpha\Gamma_1^\mu\gamma_\alpha + 
\gamma^\alpha\Gamma_1^\mu\Gamma_{1\alpha} \right) \nonumber \\
&& +4m\Big[6\Big(\Gamma_1^\alpha \{\Gamma_1^\mu,\gamma_\mu\}\gamma_\alpha + 
\gamma^\alpha \{\Gamma_1^\mu,\gamma_\mu\}\Gamma_{1\alpha} + \gamma^\alpha \Gamma_1^\mu\Gamma_{1\mu}\gamma_\alpha \Big)
- 24 \Gamma_1^\alpha\Gamma_{1\alpha} - \gamma^\alpha \{\Gamma_1^\mu,\gamma^\nu\}\{\Gamma_1^\rho,\gamma^\sigma\}\gamma_\alpha
\eta_{(\mu\nu}\eta_{\rho\sigma)} \Big]  \Big]
\end{eqnarray}
and
\begin{eqnarray}
&&i\Sigma^{(2)}_{6-8}(p)=\frac{q^2}{48}I_{log}(\lambda^2) \kappa^{\theta\alpha\tau\beta} \Big[ 
8\Gamma_{1\beta}\left[3(\cancel{p}+m)\eta_{\theta \tau}-2\gamma^\rho p_{(\theta}\eta_{\tau\rho)} \right]\gamma_\alpha
+ 8\gamma_\beta\left[3(\cancel{p}+m)\eta_{\theta \tau}-2\gamma^\rho p_{(\theta}\eta_{\tau\rho)} \right]\Gamma_{1\alpha} \nonumber \\
&& + \gamma_\beta \gamma^\rho \{\Gamma_1^\mu,\gamma^\nu\}\gamma_\alpha 
\left[3 p^\sigma \eta_{(\mu\nu}\eta_{\theta\tau}\eta_{\rho\sigma)}+4p_{\theta}\eta_{(\tau\mu}\eta_{\nu\rho)}
+4p_\tau \eta_{(\theta\mu}\eta_{\nu\rho)} \right] + 4m\gamma_\beta \{\Gamma_1^\mu,\gamma^\nu\}\gamma_\alpha
\eta_{(\mu\nu}\eta_{\theta\tau)}
\Big],	
\end{eqnarray}
in which $\{A,B\}$ stands for the anticommutation of the matrices $A$ and $B$. The result for the fifth graph is the one of equation (\ref{kappa2}). In the first-order, we have
\begin{eqnarray}
&&i\Sigma^{(1)}(p)=\frac{q^2}{12}\Big[p_{(\mu}\eta_{\nu\rho)}\gamma^\alpha \gamma^\rho \{\Gamma_1^\mu,\gamma^\nu\}\gamma_\alpha
-6p_\mu\left(-\gamma^\mu\gamma_\alpha\Gamma_1^\alpha-\Gamma_1^\alpha\gamma_\alpha\gamma^\mu 
+ \gamma^\alpha \{\Gamma_1^\mu,\gamma_\alpha\} \right) \nonumber \\
&& +3m\left[\gamma^\alpha \{\Gamma_1^\mu,\gamma_\mu\}\gamma_\alpha - 4 \{\Gamma_1^\mu,\gamma_\mu\} \right]
+ 16 \gamma_\alpha p_\beta \kappa_\rho\,^{\beta\rho\alpha}  \Big].
\end{eqnarray}
Collecting the results above, we obtain, for the coefficient $A_\Gamma^\mu$, up to the second order in the parameters
\begin{equation}
A_\Gamma^\mu=(\gamma^\mu + \rho_1^\mu + \rho_2^\mu)q^2,
\end{equation}
with
\begin{eqnarray}
\rho_1^\mu=\frac{1}{12}\Big[\delta^\mu_{(\theta}\eta_{\nu\rho)}\gamma^\alpha \gamma^\rho \{\Gamma_1^\mu,\gamma^\nu\}\gamma_\alpha
+ 6\Big(\gamma^\mu\gamma_\alpha\Gamma_1^\alpha + \Gamma_1^\alpha\gamma_\alpha\gamma^\mu 
- \gamma^\alpha \{\Gamma_1^\mu,\gamma_\alpha\} \Big) + 16 \gamma_\alpha \kappa_\rho\,^{\mu\rho\alpha}  \Big]
\label{rho1}
\end{eqnarray}
and
\begin{eqnarray}
&&\rho_2^\mu=\frac{1}{96}\Big[8\delta^\mu_{(\beta}\eta_{\nu\rho)}\Big(\Gamma_1^\alpha \gamma^\rho 
\{\Gamma_1^\mu,\gamma^\nu\}\gamma_\alpha + \gamma_\alpha \gamma^\rho \{\Gamma_1^\mu,\gamma^\nu\} \Gamma_1^\alpha 
+ \gamma^\alpha \Gamma_1^\rho \{\Gamma_1^\mu,\gamma^\nu\} \gamma_\alpha 
+ \gamma^\alpha \gamma^\rho \Gamma_1^\mu \Gamma_1^\nu \gamma_\alpha \Big) + \nonumber \\
&& - \delta^\mu_{(\beta}\eta_{\nu\rho}\eta_{\sigma\theta)}\gamma^\alpha \gamma^\theta \{\Gamma_1^\mu,\gamma^\nu\}\{\Gamma_1^\rho,\gamma^\sigma\}\gamma_\alpha 
- 48 \left(\Gamma_1^\alpha\gamma^\mu\Gamma_{1\alpha} + \Gamma_1^\alpha\Gamma_1^\mu\gamma_\alpha + 
\gamma^\alpha\Gamma_1^\mu\Gamma_{1\alpha} \right) + \nonumber \\
&& + 2\kappa^{\theta\alpha\tau\beta} \Big[ 
8\Big(3(\Gamma_{1\beta}\gamma^\mu \gamma_\alpha +\gamma_\alpha \gamma^\mu \Gamma_{1\beta})
\eta_{\theta \tau}-2(\Gamma_{1\beta}\gamma^\rho \gamma_\alpha 
+ \gamma_\alpha\gamma^\rho\Gamma_{1\beta})\delta^\mu_{(\theta}\eta_{\tau\rho)} \Big) \nonumber \\
&& + \gamma_\beta \gamma^\rho \{\Gamma_1^\lambda,\gamma^\nu\}\gamma_\alpha 
\Big(3 \eta^{\mu\sigma} \eta_{(\lambda\nu}\eta_{\theta\tau}\eta_{\rho\sigma)} + 4\delta^\mu_{\theta}\eta_{(\tau\lambda}\eta_{\nu\rho)}
+4\delta^\mu_\tau \eta_{(\theta\lambda}\eta_{\nu\rho)} \Big)\nonumber \\
&& - 4\kappa^{\gamma\sigma\tau\delta}\kappa^{\theta\nu\lambda}_{\,\,\,\,\,\,\,\,\,\delta} 
\Big(\eta_{\sigma\nu}\delta^\mu_{(\lambda}\gamma_\theta\eta_{\tau\sigma)}+3\left(2\gamma_\sigma\delta^\mu_\nu -\gamma^\mu\eta_{\sigma\nu}\right)\eta_{(\theta\lambda}\eta_{\tau\gamma)}\Big)
\Big]
\Big]
\label{rho2}.
\end{eqnarray}

The calculations for $-iq\Lambda^\mu$ are so that $B_\Gamma^\mu=-q A_\Gamma^\mu$ and $(\delta)^\mu_\nu=(\bar{\delta})^\mu_\nu$, as required by the Ward Identity of equation (\ref{Ward identity}). In the next section, we carry out the study of the beta-functions for the model, based in the above results.

\section{Beta functions}
In this section, we intend to study the beta-functions associated with the Lorentz-violating parameters. First, we present the general procedure to be adopted. The relationship between a bare (left-hand side) and a renormalized (right-hand side) coupling constant in the minimal extended QED is written as follows:
\begin{equation}
x^{J}_{B} = (Z_x)_{J'}^{J} x^{J'},    
\end{equation} 
in which $x^{J}_{B}$ and $(Z_x)_{J'}^{J}$ stand for 
\begin{equation}
x^{J}_{B} = \left\{ c^ {\mu\nu}_{B}\,\,\,;\,\,\,d^ {\mu\nu}_{B}\,\,\,;\,\,\,e^{\mu}_{B}\,\,\,;\,\,\,f^{\mu}_{B}\,\,\,;\,\,\,g^{\mu\nu\rho}_{B} \right\} \label{couplings}
\end{equation}
and
\begin{equation}
(Z_x)_{J'}^{J}x^{J'} = \left\{\left(Z_{c}\right)_{\alpha\beta}^{\nu\mu}c^{\alpha\beta}\,\,\,;\,\,\, \left(Z_{d}\right)_{\alpha\beta}^{\nu\mu}d^{\alpha\beta}\,\,\,;\,\,\,\left(Z_{e}\right)_{\alpha}^{\nu}e^{\alpha}\,\,\, ;\,\,\,\left(Z_{f}\right)_{\alpha}^{\nu}f^{\alpha}\,\,\, ;\,\,\, \left(Z_{g}\right)_{\alpha\beta\gamma}^{\lambda\nu\mu}g^{\alpha\beta\gamma}\right\}\label{BareRenorm}. 
\end{equation}
Above, $J$ and $J'$ represent the free Lorentz indices associated with each one of the elements in the sets. The beta-function for the couplings are defined by
\begin{equation}
\beta_{x}^{J}=2\lambda^{2}
\frac{d x^{J} }{d(\lambda^{2})},
\label{defBetac}
\end{equation}
where $\lambda$ is the renomalization group scale, which is represented by the argument of the basic divergence $I_{log}(\lambda^2)$ defined in section II. The renormalization constant $(Z_x)_{J'}^{J}$ can be obtained, using the definitions of equation (\ref{twoPoint}), by
\begin{equation}
Z_x Z_2 x = x + iA_x I_{log}(\lambda^2),
\end{equation}
with
\begin{equation}
Z_2= 1 + iA_\psi I_{log}(\lambda^2),
\end{equation}
in which we omit the $J$ indices for simplicity. The one-loop coefficients $A_x$ are represented as series in the Lorentz-breaking parameters. We write
\begin{equation}
A_\psi=(\alpha_0+\alpha_1+\alpha_2+\cdots)q^2 \,\,\,\, \mbox{and} \,\,\,\, 
A_x=\left\{\rho_{x0}+\rho_{x1}+\rho_{x2}+\cdots\right\}q^2,
\end{equation}
in which, $\alpha_i$ and $\rho_{xi}$ are of $i$-th order in the Lorentz-violating parameters. So, we have
\begin{eqnarray}
&&x_B=Z_x  x = \left\{1 - iA_\psi I_{log}(\lambda^2)\right\}\left\{x + iA_x I_{log}(\lambda^2)\right\} \nonumber \\
&&=x + i\left\{\left[\rho_{x1}-x\alpha_0\right]+\left[\rho_{x2}-x\alpha_1\right]\right\}q^2 I_{log}(\lambda^2)+\cdots
\label{bare}
\end{eqnarray}
up to the order $q^2x^2$. We then apply $2\lambda^2 \frac{d}{d\lambda^2}$ in both sides of the equation above to find
\begin{equation}
\beta_x^J=-\frac{1}{8\pi^2}\left\{(\rho_{x1}^J-x^J\alpha_0)+(\rho_{x2}^J-x^J\alpha_1)\right\}.
\label{beta}
\end{equation}

Let us then calculate the specific results for the beta-functions. In the last section, we obtained the expressions displayed in equations (\ref{apsi})-(\ref{as}), up to second order in the Lorentz-violating parameters, for the coefficients $A_x^J$. It is to be noted that, among the Lorentz-breaking parameters, only $c^{\mu\nu}$ has contributions of second-order to the corresponding $A_c^{\mu\nu}$. Consequently, if the crossed terms are not considered, only the beta-function for $c^{\mu\nu}$ will have a non-zero second-order contribution in the parameters. For the beta-functions, we obtain, with the use of equation (\ref{beta}):
\begin{eqnarray}
&&\beta_e^\mu=\beta_f^\mu=\beta_h^\mu=0, \,\,\,\beta_d^{\mu\nu}=\frac{q^2}{3\pi^2}d^{\mu\nu}, \nonumber \\
&&\beta_c^{\mu\nu}=\frac{q^2}{12\pi^2}\left\{2\left(2c^{\mu\nu}-\kappa^{\mu\theta\nu}_{\,\,\,\,\,\,\,\,\theta}\right)
+\left(3c^{\mu\sigma}c_\sigma^\nu+3d^{\mu\sigma}d_\sigma^\nu-2e^\mu e^\nu-2f^\mu f^\nu-2h^\mu h^\nu 
- \frac 12 G^{\mu\nu}(\kappa) \right)\right\}.
\end{eqnarray}

The first-order result is known from \cite{Alan2}. The difference in our result for $\beta_g^{\mu\nu\alpha}$ is due to the fact that we have considered a case of the completely antisymmetric tensor $g^{\mu\nu\alpha}=\varepsilon^{\mu\nu\alpha\beta}h_\beta$. 

The results above for the beta-functions do not take crossed terms into account. An alternative to obtain the total second-order value for the beta-function is to use the approach developed in subsection \ref{general} and calculate
\begin{equation}
\beta_\Gamma^\mu=2\lambda^2 \frac{d\Gamma^\mu}{d\lambda^2}.
\end{equation}
We start from the following equation:
\begin{eqnarray}
&&\Gamma^\mu_B=(Z_\Gamma)^\mu_\nu  \Gamma^\nu = \left\{1 - iA_\psi I_{log}(\lambda^2)\right\}\left\{\Gamma^\mu + iA_\Gamma^\mu I_{log}(\lambda^2)\right\} \nonumber \\
&&=\Gamma^\mu+i(A_\Gamma^\mu-A_\psi \Gamma^\mu)I_{log}(\lambda^2) \nonumber \\
&&=\Gamma^\mu + i\left\{\gamma^\mu+(\rho_1^\mu-\Gamma^\mu\alpha_0)+(\rho_2^\mu-\Gamma^\mu \alpha_1) \right\}q^2 I_{log}(\lambda^2),
\end{eqnarray}
with $\rho_1^\mu$ and $\rho_2^\mu$ given by equations (\ref{rho1}) and (\ref{rho2}), respectively. If now we apply the operator $2\lambda^2 d/d\lambda^2$ in both sides of the above equation, and take in consideration that $\alpha_0=1$ and $\alpha_1=0$, we get
\begin{equation}
\beta_{\Gamma}^\mu=-\frac{1}{8\pi^2}\left\{(\rho_1^\mu-\Gamma^\mu) + \rho_2^\mu \right\}.
\end{equation}
With the equation above, we can get all the crossed terms. Noting that $A^\mu_{\Gamma}=A^\mu_{\Gamma_1}+A_\psi \gamma^\mu$, such that
\begin{equation}
\rho_i^\mu=\tilde{\rho}_i^\mu+\alpha_i \gamma^\mu,
\end{equation}
with $\tilde{\rho}_i^\mu$ the coefficients of the expansion of $A_{\Gamma_1}$, all the beta-functions discussed before can be recovered by means of the following decomposition: 
\begin{equation}
\beta_{\Gamma}^\mu - \frac{1}{8\pi^2}\alpha_2 \gamma^\mu=\beta_{e}^{\mu}+i \beta^\mu \gamma_5  +  \beta_{c}^{\mu\nu}\gamma_{\nu} + 
\beta_{d}^{\mu\nu}\gamma_5\gamma_{\nu} + \frac 12 \beta_{g}^{\alpha\beta\mu}\sigma_{\alpha\beta}.
\end{equation}

The second-order result brings us new conclusions. The inclusion of some of the Lorentz-breaking terms is linked to the presence of others. For example, a model with terms involving $e^\mu$, $f^\mu$, $g^{\mu\nu\alpha}$ or $\kappa^{\mu\nu\alpha\beta}$ and without the $c^{\mu\nu}$ term is inconsistent from the renormalizability viewpoint, since the $c^{\mu\nu}$ term receives divergent contributions from other terms. This conclusion is also valid for the Lorentz-violating CPT-even term in the photon sector: it is needed if one includes at least one of the terms in $e^\mu$, $f^\mu$, $g^{\mu\nu\alpha}$ and $c^{\mu\nu}$. As the explicit results for the crossed terms shown in the Appendix B make evident, it all the parameters receive divergent contributions at second-order.

\section{Conclusion}
We carried out a study of the one-loop corrections to the minimal extended QED up to second-order in the Lorentz-violating parameters. At first glance, such an investigation might seem unnecessary, considering the very low upper limit for the magnitude of such background vectors imposed by experimental results. This is the reason why almost all articles devoted to the Standard Model Extension are focused on the study of first-order quantum corrections effects. However, as pointed out in the introduction, there are some subtleties, which were clarified in the present paper, that, in a way, suggest some care in  the understanding of the model as a whole.

First of all, the first order corrections in some of these parameters are null. In these cases, it is advisable to take a look at the lowest-order non null correction. In a previous paper \cite{sec-order}, the second-order one-loop corrections to the photon sector of the extended QED where studied, including finite parts. It was shown that some of the parameters do not cause quantum inductions in the photon sector at all, whereas others contribute depending on the order of calculation. In that study, it is instructive the cases of the parameters $e^\mu$ and $f^\mu$, which do not contribute in the first order calculation. At the second order, however, $e^\mu$ and $f^\mu$ furnish divergent amplitudes that affect both {\bf the} Maxwell and the Lorentz-violating CPT-even (also called here aether) terms.

In this paper, we went deeper in the analysis of the model up to second order contributions. It is meaningful that various of the background tensors are linked to each other. When someone selects a single term to formulate a simplified Lorentz-breaking extension of QED, for example, it will likely also be necessary to include, for consistency, one or more other terms. And that view can be limited by an analysis that takes into account only first-order corrections.

The second-order results obtained in the present paper enforce the previous conclusions. As we already noted above, a model with terms proportional to $e^\mu$, $f^\mu$ or $g^{\mu\nu\alpha}$ but without that one proportional to $c^{\mu\nu}$ is inconsistent, since the $c^{\mu\nu}$ term receives divergent contributions depending on $e^\mu$, $f^\mu$ or $g^{\mu\nu\alpha}$. This conclusion is also valid for the Lorentz-violating CPT-even term in the photon sector: it is needed if one includes at least one of the terms involving $e^\mu$, $f^\mu$, $g^{\mu\nu\alpha}$ and $c^{\mu\nu}$. Our calculation of the beta-functions also demonstrates how these parameters are connected. Besides, the preliminary calculations of the crossed terms show that even more connections can be found, mainly if higher-order corrections are considered. A more complete study of these mixed terms is left for future studies.

\section*{Appendix A: Feynman integrals for the vertex quantum corrections}
\begin{eqnarray}
&&\Lambda^{(2)\mu}_1=-q^3\int\frac{d^4k}{(2\pi)^4}\,\frac{\gamma^\alpha(\cancel{p}'-\cancel{k}+m)\Gamma_1^\mu
(\cancel{p}-\cancel{k}+m)\Gamma_{1\alpha}}
{k^2\left[(p'-k)^2-m^2\right]\left[(p-k)^2-m^2\right]};\\
&&\Lambda^{(2)\mu}_2=-q^3\int\frac{d^4k}{(2\pi)^4}\,\frac{\Gamma_1^\alpha(\cancel{p}'-\cancel{k}+m)\Gamma_1^\mu
	(\cancel{p}-\cancel{k}+m)\gamma_{\alpha}}
{k^2\left[(p'-k)^2-m^2\right]\left[(p-k)^2-m^2\right]};\\
&&\Lambda^{(2)\mu}_3=q^3\int\frac{d^4k}{(2\pi)^4}\,\frac{(p-k)_\rho \gamma^\alpha(\cancel{p}'-\cancel{k}+m)\Gamma_1^\mu
	(\cancel{p}-\cancel{k}+m)\Gamma_1^\rho(\cancel{p}-\cancel{k}+m)\gamma_{\alpha}}
{k^2\left[(p'-k)^2-m^2\right]\left[(p-k)^2-m^2\right]^2};\\
&&\Lambda^{(2)\mu}_4=q^3\int\frac{d^4k}{(2\pi)^4}\,\frac{(p'-k)_\rho \gamma^\alpha(\cancel{p}'-\cancel{k}+m)
	\Gamma_1^\rho (\cancel{p}'-\cancel{k}+m) \Gamma_1^\mu (\cancel{p}-\cancel{k}+m)\gamma_{\alpha}}
{k^2\left[(p'-k)^2-m^2\right]^2\left[(p-k)^2-m^2\right]};\\
&&\Lambda^{(2)\mu}_5=-q^3\int\frac{d^4k}{(2\pi)^4}\,\frac{(p'-k)_\rho (p-k)_\sigma \gamma^\alpha(\cancel{p}'-\cancel{k}+m)
	\Gamma_1^\rho (\cancel{p}'-\cancel{k}+m) \gamma^\mu (\cancel{p}-\cancel{k}+m)
	\Gamma_1^\sigma (\cancel{p}-\cancel{k}+m)\gamma_{\alpha}}
{k^2\left[(p'-k)^2-m^2\right]^2\left[(p-k)^2-m^2\right]^2};\\
&&\Lambda^{(2)\mu}_6=q^3\int\frac{d^4k}{(2\pi)^4}\,\frac{(p-k)_\rho \Gamma_1^\alpha(\cancel{p}'-\cancel{k}+m)
	 \gamma^\mu (\cancel{p}-\cancel{k}+m)\Gamma_1^\rho (\cancel{p}-\cancel{k}+m)\gamma_{\alpha}}
{k^2\left[(p'-k)^2-m^2\right]\left[(p-k)^2-m^2\right]^2};\\
&&\Lambda^{(2)\mu}_7=q^3\int\frac{d^4k}{(2\pi)^4}\,\frac{(p'-k)_\rho \gamma^\alpha(\cancel{p}'-\cancel{k}+m)
	\Gamma_1^\rho (\cancel{p}'-\cancel{k}+m)
	\gamma^\mu (\cancel{p}-\cancel{k}+m)\Gamma_{1\alpha}}
{k^2\left[(p'-k)^2-m^2\right]^2\left[(p-k)^2-m^2\right]};\\
&&\Lambda^{(2)\mu}_8=q^3\int\frac{d^4k}{(2\pi)^4}\,\frac{(p'-k)_\rho \Gamma_1^\alpha(\cancel{p}'-\cancel{k}+m)
	\Gamma_1^\rho (\cancel{p}'-\cancel{k}+m)
	\gamma^\mu (\cancel{p}-\cancel{k}+m)\gamma_{\alpha}}
{k^2\left[(p'-k)^2-m^2\right]^2\left[(p-k)^2-m^2\right]};\\
&&\Lambda^{(2)\mu}_9=q^3\int\frac{d^4k}{(2\pi)^4}\,\frac{(p-k)_\rho \gamma^\alpha(\cancel{p}'-\cancel{k}+m)
	\gamma^\mu (\cancel{p}-\cancel{k}+m)\Gamma_1^\rho (\cancel{p}-\cancel{k}+m)\Gamma_{1\alpha}}
{k^2\left[(p'-k)^2-m^2\right]\left[(p-k)^2-m^2\right]^2};\\
&&\Lambda^{(2)\mu}_{10}=-q^3\int\frac{d^4k}{(2\pi)^4}\,\frac{\Gamma_1^\alpha(\cancel{p}'-\cancel{k}+m)
	\gamma^\mu (\cancel{p}-\cancel{k}+m)\Gamma_{1\alpha}} {k^2\left[(p'-k)^2-m^2\right]\left[(p-k)^2-m^2\right]};\\
&&\Lambda^{(2)\mu}_{11}=-q^3\int\frac{d^4k}{(2\pi)^4}\,\frac{(p-k)_\rho (p-k)_\sigma \gamma^\alpha(\cancel{p}'-\cancel{k}+m)
	 \gamma^\mu (\cancel{p}-\cancel{k}+m)\Gamma_1^\rho (\cancel{p}-\cancel{k}+m)
	\Gamma_1^\sigma (\cancel{p}-\cancel{k}+m)\gamma_{\alpha}}
{k^2\left[(p'-k)^2-m^2\right]\left[(p-k)^2-m^2\right]^3};\\
&&\Lambda^{(2)\mu}_{12}=-q^3\int\frac{d^4k}{(2\pi)^4}\,\frac{(p'-k)_\rho (p'-k)_\sigma \gamma^\alpha(\cancel{p}'-\cancel{k}+m)
	\Gamma_1^\rho (\cancel{p}'-\cancel{k}+m) \Gamma_1^\sigma (\cancel{p}'-\cancel{k}+m)
	\gamma^\mu (\cancel{p}-\cancel{k}+m)\gamma_{\alpha}}
{k^2\left[(p'-k)^2-m^2\right]^3\left[(p-k)^2-m^2\right]}; \\
&&\Lambda^{(2)\mu}_{13}=-4q^3 \kappa^{\gamma\sigma\theta\delta}\kappa^{\theta\nu\lambda}_{\,\,\,\,\,\,\,\,\,\delta}
\int\frac{d^4k}{(2\pi)^4}\,\frac{\gamma_\sigma (\cancel{p}'-\cancel{k}+m)
	\gamma^\mu (\cancel{p}-\cancel{k}+m)\gamma_{\nu}k_\theta k_\lambda k_\tau k_\gamma}
{k^6\left[(p'-k)^2-m^2\right]\left[(p-k)^2-m^2\right]};\\
&&\Lambda^{(2)\mu}_{14}=2q^3 \kappa^{\lambda\alpha\nu\beta}\int\frac{d^4k}{(2\pi)^4}\,\frac{\gamma_{\nu}(\cancel{p}'-\cancel{k}+m)\gamma^{\mu}
(\cancel{p}-\cancel{k}+m)\Gamma_{1\lambda}k_\alpha k_\beta}
{k^4\left[(p'-k)^2-m^2\right]\left[(p-k)^2-m^2\right]};\\
&&\Lambda^{(2)\mu}_{15}=-2q^3 \kappa^{\lambda\alpha\nu\beta}\int\frac{d^4k}{(2\pi)^4}\,\frac{(p-k)_\rho\gamma_{\nu}(\cancel{p}'-\cancel{k}+m)\gamma^{\mu} 
(\cancel{p}-\cancel{k}+m)\Gamma^\rho_{1}(\cancel{p}-\cancel{k}+m)\gamma_{\lambda}k_\alpha k_\beta}
{k^4\left[(p'-k)^2-m^2\right]\left[(p-k)-m^2\right]^2};\\
&&\Lambda^{(2)\mu}_{16}=-2q^3 \kappa^{\lambda\alpha\nu\beta}\int\frac{d^4k}{(2\pi)^4}\,\frac{(p'-k)_\rho\gamma_{\nu}(\cancel{p}'-\cancel{k}+m)\Gamma^\rho_{1} 
	(\cancel{p}'-\cancel{k}+m)\gamma^{\mu}(\cancel{p}-\cancel{k}+m)\gamma_{\lambda}k_\alpha k_\beta}
{k^4\left[(p'-k)^2-m^2\right]^2\left[(p-k)-m^2\right]};\\
&&\Lambda^{(2)\mu}_{17}=2q^3 \kappa^{\lambda\alpha\nu\beta}\int\frac{d^4k}{(2\pi)^4}\,\frac{\gamma_{\nu}(\cancel{p}'-\cancel{k}+m)\Gamma_1^{\mu}
(\cancel{p}-\cancel{k}+m)\gamma_\lambda k_\alpha k_\beta}
{k^4\left[(p'-k)^2-m^2\right]\left[(p-k)^2-m^2\right]};\\
&&\Lambda^{(2)\mu}_{18}=2q^3 \kappa^{\lambda\alpha\nu\beta}\int\frac{d^4k}{(2\pi)^4}\,\frac{\Gamma_{1\nu}(\cancel{p}'-\cancel{k}+m)\gamma^{\mu}
(\cancel{p}-\cancel{k}+m)\gamma_\lambda k_\alpha k_\beta}
{k^4\left[(p'-k)^2-m^2\right]\left[(p-k)^2-m^2\right]}.
\end{eqnarray}

\section*{Appendix B: Some crossed terms contributions for two-point functions}

Here, we present, for completeness, some results for the crossed terms of the two-point function (amplitudes depicted in equation ($30$)):
\begin{equation}
i\Sigma^{(2ef)}= -3 q^2 i m I_{log}(\lambda^{2}) (e \cdot f) \gamma_{5},
\end{equation}
\begin{equation}
i\Sigma^{(2eh)}= -\frac{1}{3}iq^{2}I_{log}(\lambda^{2})\Big\{2 \cancel{p} \epsilon^{\alpha\mu\nu\beta}\gamma_{\mu}\gamma_{\nu}e_{\alpha}h_{\beta} + \cancel{e} \epsilon^{\alpha\mu\nu\beta}\gamma_{\mu}\gamma_{\nu}p_{\alpha}h_{\beta}-\epsilon^{\alpha\mu\nu\beta}\gamma_{\mu}p_{\alpha}e_{\nu}h_{\beta}-\epsilon^{\mu\nu\alpha\beta}\gamma_{\mu}\gamma_{\nu}\gamma_{\alpha}h_{\beta}(e \cdot p)\Big\},
\end{equation}
\begin{eqnarray}
i\Sigma^{(2fh)}&=&-\frac{1}{8}q^{2}I_{log}(\lambda^{2})\epsilon^{\mu\nu\alpha\beta}h_{\beta}\gamma_{5}\Bigg\{\frac{1}{3}\gamma_{\mu}\gamma_{\nu}\gamma_{\alpha}\cancel{e}\cancel{f}-4m\gamma_{\mu}\gamma_{\alpha}f_{\nu}+\frac{1}{3}\cancel{p}\gamma_{\mu}\gamma_{\nu}f_{\alpha}-6\gamma_{\mu}f_{\nu}p_{\alpha}-\cancel{f}\gamma_{\nu}\gamma_{\alpha}p_{\mu}\nonumber \\&&-\frac{1}{3}\gamma_{\mu}\gamma_{\nu}\gamma_{\alpha}(f \cdot p)\Bigg\},
\end{eqnarray}
\begin{equation}
i\Sigma^{(2ce)}=-\frac{1}{3}q^{2}I_{log}(\lambda^{2})c^{\mu\nu}\Big\{8 e_{\mu}p_{\nu}-m e_{\mu}\gamma_{\nu}\Big\},
\end{equation}
\begin{equation}
i\Sigma^{(2cf)}=-\frac{8}{3} i q^{2} \gamma_{5} c^{\mu\nu}f_{\mu}p_{\nu},
\end{equation}
\begin{equation}
i\Sigma^{(2de)}=\frac{1}{3}q^2I_{log}(\lambda^{2})d^{\mu\nu}\gamma_{5}\Bigg\{2 p_{\mu}e_{\nu}-5 m \gamma_{\mu}e_{\nu}+p_{\mu}\cancel{e}\gamma_{\nu}-\frac{1}{2}e_{\mu}\cancel{e}\gamma_{\nu}\Bigg\},
\end{equation}
\begin{equation}
i\Sigma^{(2df)}=\frac{1}{3} i q^{2} I_{log}(\lambda^{2})d^{\mu\nu}\Bigg\{2p_{\mu}f_{\nu}+p_{\mu}\cancel{f}\gamma_{\nu}-3f_{\nu}\cancel{p}\gamma_{\mu}\Bigg\},
\end{equation}
\begin{equation}
i\Sigma^{(2cd)}=-I_{log}(\lambda^{2})\gamma_{5}\Big\{3mc^{\mu\nu}d_{\mu\nu}+2c^{\mu\alpha}d_{\mu}^{\,\,\,\,\nu}p_{\alpha}\gamma_{\nu}+c^{\mu\nu}d_{\alpha\mu}p^{\alpha}\gamma_{\nu}-\frac{11}{12}c^{\mu\nu}d_{\mu\nu}\cancel{p}-\frac{1}{6}c^{\mu\nu}d_{\alpha\nu}\cancel{p}\gamma_{\mu}\gamma^{\alpha}\Big\},
\end{equation}
\begin{equation}
i\Sigma^{(2e\kappa)}=q^{2}I_{log}(\lambda^{2})\kappa^{\mu\nu\alpha\beta}\gamma_{\alpha}\gamma_{\nu}e_{\mu}p_{\beta},
\end{equation}
\begin{equation}
i\Sigma^{(2f\kappa)}=\frac{1}{3} i q^{2} I_{log}(\lambda^{2}) \kappa^{\mu\alpha\nu\beta}\gamma_{\mu}\gamma_{\nu}f_{\alpha}p_{\beta},
\end{equation}
\begin{equation}
i\Sigma^{(2h\kappa)}=\frac{1}{2}i q^2 I_{log}(\lambda^{2})\epsilon^{\beta\alpha\nu\sigma}\gamma^{\mu}\Bigg\{\Big(mh_{\beta}\kappa_{\mu\alpha\nu\sigma}+\gamma_{\sigma}\kappa_{\lambda\nu\alpha\mu}h_{\beta}p^{\lambda}\Big) +\frac{1}{4}i\gamma^{\mu}\gamma^{\nu}\epsilon^{\lambda\alpha\beta\sigma}\Big(mh_{\lambda}\gamma_{\sigma}\kappa_{\mu\alpha\nu\beta}+h_{\lambda}p_{\sigma}\kappa_{\mu\alpha\nu\beta}\Big)\Bigg\},
\end{equation}
\begin{equation}
i\Sigma^{(2c\kappa)}=\frac{1}{3}q^2 I_{log}(\lambda^{2})\Bigg\{m c_{\alpha\beta}\gamma_{\mu}\gamma_{\nu}\kappa^{\mu\alpha\nu\beta} + c_{\alpha\beta}\cancel{p} \kappa^{\mu\alpha\nu\beta} + 5 c_{\mu\alpha}\gamma_{\nu}\kappa^{\alpha\beta\mu\nu}p_{\beta}+3c_{\alpha\mu}\gamma_{\nu}\kappa^{\mu\beta\nu\alpha}p_{\beta} - c_{\alpha\mu}\gamma_{\nu}\kappa^{\nu\beta\mu\alpha}p_{\beta}\Bigg\},
\end{equation}
\begin{equation}
i\Sigma^{(2d\kappa)}=\frac{1}{3}q^{2}I_{log}(\lambda^{2})\Bigg\{2 d_{\mu\nu}\gamma_{\alpha}\gamma_{5}\kappa^{\nu\alpha\mu\beta}p_{\beta} + m d_{\alpha\mu}\gamma_{\beta}\gamma_{\nu}\gamma_{5}\kappa^{\mu\alpha\nu\beta}+\frac{1}{2}d_{\alpha\beta}\cancel{p} \gamma_{\mu}\gamma_{\nu}\gamma_{5}\kappa^{\mu\beta\nu\alpha}-d_{\mu\nu}\gamma_{\alpha}\gamma_{5}\kappa^{\mu\nu\alpha\beta}p_{\beta}\Bigg\}
\end{equation}

 {\bf Acknowledgments.} A.P.B.S acknowledges CNPq by financial support. The work by A. Yu. P. has been partially supported by the
CNPq project No. 301562/2019-9.

\end{document}